\documentclass[aps,twocolumn,nofootinbib,superscriptaddress]{revtex4-1}

\pdfoutput=1

\usepackage{slashed}

\usepackage[normalem]{ulem}

\usepackage{color}

\usepackage{epsfig}
\usepackage{epstopdf}
\usepackage{epsf}
\input{epsf.sty}
\usepackage{graphicx,amsmath,amssymb}
\usepackage{epsfig}
\allowdisplaybreaks

\def\ei{\end{itemize}}
\def\be{\begin{equation}}
\def\ee{\end{equation}}
\newcommand{\bea}{\begin{eqnarray}}
\newcommand{\eea}{\end{eqnarray}}

\usepackage{bbm,bm,amsmath,amssymb}

\usepackage{hyperref}

\begin{document}

\title{Primordial Circular Polarization in the Cosmic Microwave Background}

\author{Stephon Alexander}
\email{stephon\_alexander@brown.edu}
\author{Evan McDonough}
\email{evan\_mcdonough@brown.edu}
\affiliation{Department of Physics, Brown University, Provide, RI.}

 \begin{abstract}
Circular (``V-mode'') polarization is expected to be vanishing in the CMB, since it is not produced in Thomson scattering.  However, considering that the conventional CMB anisotropies are generated via an early universe mechanism such as inflation or a bouncing scenario, it is possible that circular polarization could be primordially produced and survive to the surface of last scattering. We study this in detail, and find a large class of inflationary models that produce a nearly scale invariant spectrum of scalar V-mode anisotropies.  We study the inflationary production and subsequent evolution via the Boltzmann hierarchy, and show that V-mode polarization present in the CMB is suppressed by a factor of at least $10^{10^{20}}$ relative to the primordial $V$, consistent with expectation of negligible V-mode polarization from inflation.  We consider alternative possibilities for sourcing V primordially, such as the V-mode polarization induced by circularly polarized primordial gravitational waves, or producing $V$ after inflation, via new interactions at recombination.
\end{abstract}

\maketitle

%=================================================================%
\section{Introduction}
\label{sec:intro}
%=================================================================%

Circular (``V-mode'') polarization is an exciting new window into the physics of the cosmic microwave background. It is a direct probe of early universe CP-violation in the standard model sector, which is itself a necessary condition for baryogenesis \cite{Sakharov:1967dj}. On the observational front, the SPIDER experiment \cite{Nagy:2017csq} has put the first ever constraints on the large scale V-mode anisotropies $\delta V < 10^{2} \mu{\rm K}$, and more constraints are expected in the near future. However, the theory side of circular polarization, namely predictions for models of the early universe, remains largely unstudied.

The primary obstacle to observing any \emph{primordial} circular polarization, i.e.~a net polarization due to the physics of inflation (or alternatives), is the dissipation of circular polarization that occurs after inflation but before last scattering. For this reason, the primordial $V$ is expected to vanish, and the focus of much work to date (e.g.~\cite{Cooray:2002nm,De:2014qza,Montero-Camacho:2018vgs,Mohammadi:2013dea,Motie:2011az,Sawyer:2014maa,*Sawyer:2012gn,Sadegh:2017rnr,Kamionkowski:2018syl,Giovannini:2009ru,*Giovannini:2010ar,Ejlli:2018ucq,*Ejlli:2016avx,*Ejlli:2016avx}) has been on `conventional' sources of circular polarization, e.g. Faraday conversion \cite{Cooray:2002nm}, which in many cases leads to a negligible large-scale CMB signal \cite{Montero-Camacho:2018vgs} ($\lesssim 10^{-14}$ K).  However, no attempt has been made to quantify this expectation, and with this in mind, in this work we consider in detail the production and evolution of primordial circular polarization. 

There exist general mechanisms for the primordial generation of $V$, and we will provide two examples in the context of inflationary cosmology\footnote{Circular polarization undoubtedly arises in alternatives to inflation, and may in fact a useful discriminator of early universe scenarios, but in this work we focus on inflation.}. In particular, it was shown by the authors in \cite{Alexander:2017bxe} that axion inflation coupled to fermions and gauge fields leads to primordial V-mode polarization, from the direct production of particles during inflation. In this paper we demonstrate that a scale-invariant spectrum of $V$ anisotropies can arise via a minimal modification of these models.

The evolution of the inflationary $V$ to last scattering requires a careful study of the Boltzmann hierarchy for $V$. The collision term for scalar brightness temperature perturbations $V_{\ell}$ is given by the $\ell=1$ the anisotropies $V_1$, which can in turn be computed by solving the Boltzmann hierarchy for $V_{0}$ and $V_1$. Upon performing this analysis, we find that the power spectrum $C^{\ell} _{VV}$ is suppressed relative to the inflationary $V$ by a factor  $\gtrsim 10^{10^{20}}$, which can be qualitatively understood as a consequence of  the high conductivity of the universe during the radiation dominated era.

While this result is a negative one, it has an important implication: any observed non-zero $V$ \emph{must} be due to new physics occurring \emph{after} inflation (or an alternative). This statement is of course subject to caveats, and we will briefly discuss  one such loop-hole: the sourcing of $V$ via primordial circularly polarized gravitational waves\footnote{A similar claim was reached in a very recent work \cite{Inomata:2018rin}, which appeared during the final stages of preparation of this paper.}.

The structure of this note is as follows: we first provide an overview of conventions and notation, and then proceed in Section III to study the production of $V$ during inflation. In section IV we solve the Boltzmann hierarchy and estimate the $V$ present in the CMB, and find a large suppression.  We close in section V with a discussion of other sources of $V$ and directions for future work.

%=================================================================%
\section{The Stokes V parameter and V-mode Anisotropies}
\label{sec:inflation}
%=================================================================%

We begin by defining the notation and relevant quantities for the study of circular polarization. For an electromagnetic wave $\vec{\epsilon}$ propagating in the $\hat{z}$-direction, the Stokes parameters are defined as,
\bea
&&I= \frac{1}{a^2}
\left( |E_x|^2 + |E_y|^2 \right) \; , \;  V =\frac{1}{a^2}\left( E_x ^* E_y - E_y ^* E_x\right) \\
 &&Q=  \frac{1}{a^2}\left( |E_x|^2 - |E_y|^2 \right) \; ,\; U= \frac{1}{a^2}\left( E_x^* E_y + E_y^* E_x\right).
 \eea
The circular polarization $V$ can alternately be written as the phase difference $\Delta\phi$ between its $x$ and $y$ components,
\be
 V=\frac{2}{a^2}\epsilon_x\epsilon_y\sin\Delta\phi ,
 \ee
while $U \propto \cos \Delta \phi$.  The nature of $V$ as circular polarization is made manifest by a change of basis to $\{x_+,x_-\}$ coordinates, $\sqrt{2} \hat{x}_+ = \hat{x} + i \hat{y}$, $\sqrt{2} \hat{x}_- = \hat{x} - i \hat{y}$. In this basis, in an FRW spacetime,
\begin{equation}
V = \frac{1}{a^4} \left( |E_+  |^2 - |E_- |^2\right) ,
\end{equation}
and hence $V$ is the difference in amplitude of the two circular polarization states.

The anisotropies in $V$\footnote{Similar to linear polarization, anisotropies in $V$ can be expressed as a fractional temperature fluctuation  $\Theta_V$ via the rescaling, $\Theta_V \equiv V_T/T \equiv V/I$, where $V_T$ is the brightness temperature perturbation of the circular polarization,  and $T$ and $I$ are the background blackbody temperature and intensity respectively. In this work we  interchangeably use $V_{T}$, $V$, and $
\Theta_V$, and their distinction should be obvious from the context.} are expressed in terms of mulitpole moments $V_{\ell}$, which we define via the multipole decomposition as (where now $V$ denotes the dimensionless $\Theta_V$)
\be
 V(\eta,\vec{x},\vec{n})=\int\frac{{\rm d}^3k}{(2\pi)^3}\sum_{m=-2}^2 \sum_{\ell\geq \mid m \mid}(2\ell+1)V_\ell^{(m)}G_\ell^m  ,
 \ee
where,
\be 
G_\ell^m = (-i)^\ell \sqrt{\frac{4\pi}{2\ell+1}}Y_\ell^m(\hat{n})\exp(i\vec{k}\cdot\vec{x}) .
\ee 
From this we define the \emph{primordial} power spectrum $P_{V}(k)$ as
\be
\label{PVkV0}
P_{V}(k) = \frac{k^3}{2 \pi^2} |V_0|^2 ,
\ee
in analogy with temperature anisotropies. This can be parametrized in terms of an amplitude $\mathcal{A}_V$ and spectral tilt $n_V$ as
\be
P_{V}(k) = \mathcal{A}_V \left( \frac{k}{k_0}\right)^{n_V - 1} .
\ee
From this we define the circular polarization to scalar ratio,
\be
\label{eq:rV}
r_{V} \equiv \frac{P_{V}}{P_{\zeta}} ,
\ee
evaluated at the CMB pivot scale, in analog with the tensor-to-scalar ratio.

The $VV$ power spectrum coefficients $C_{\ell}$ are given by the two-point correlation function, $C_{\ell} ^{VV} = \langle V_\ell V_{\ell} \rangle $, and we quantify the degree to which the CMB is circularly polarized (on large scales) via the quantity,
\be
r_{V} ^{CMB} = \frac{C_{\ell} ^{VV}}{C_{\ell} ^{TT}} |_{\ell = \mathcal{O}(1)}.
\ee
The relation between $r_V$ and $r_{V} ^{CMB} $ is determined via solving the Boltzmann equation, and evolving $V$ from inflation to last scattering.

%=================================================================%
\section{Inflation and Primordial V-modes}
\label{sec:inflation}
%=================================================================%

We start with the simplest model of axion inflation coupled to photons via the axionic coupling $\phi F \tilde{F}$. The spectrum of primordial V-modes in this scenario was computed in \cite{Alexander:2017bxe}. The action for this is given by \cite{Anber:2009ua,Barnaby:2011vw,Barnaby:2011qe},
\begin{eqnarray}
\label{eq:axionmodel}
S = & & \int {\rm d}^4 x \sqrt{- g} \left[ \frac{M_{\rm Pl}^2}{2}R - \frac{1}{2}(\partial\phi)^2 -  V(\phi)  \right. \nonumber \\
 & & -\frac{1}{4}F_{\mu\nu}F^{\mu\nu} +   \frac{\alpha}{f} \phi F_{\mu\nu}\tilde{F}^{\mu\nu} \left. \right], 
\end{eqnarray}
where $F_{\mu \nu}$ is the gauge field strength tensor, $\tilde{F}$ is the Hodge dual of $F$, and the field $\phi$ is the inflaton. The resulting equation of motion for the $(\pm)$ helicity Fourier modes $A_{k\pm}$ is given by
\begin{equation} \label{EoMA}
\frac{d^2 {A}_{k\pm}}{d \tau^2} + \left( k^2 \pm 2k \frac{\xi}{\tau} \right) A_{k\pm} \, = \, 0 \, ,
\end{equation}
where the parameter $\xi$ is defined as,
\bea
\label{xi}
\;\;\;\;\; \xi  = \, \frac{2 \alpha \dot{\phi}}{f H} .  \\ \nonumber
\eea
The definite sign of $\dot{\phi}$ during inflation leads to exponential growth of one polarization state. After imposing vacuum initial conditions, the solution for the amplified mode is given by ,
\bea
\label{modefunctioninflation}
A_{k+}   \, &=& \, 
\frac{2^{-1/4}}{\sqrt{2k}} \left( \frac{k}{ \xi a H}\right)^{1/4} e^{\pi \xi - 4 \xi \sqrt{ k / 2 \xi a H}}  .
\eea
Thus the $(+)$ mode is amplified by a factor of $e^{\pi \xi}$. The maximal amplification occurs for modes with $k\sim \xi a H$, while the amplification falls off exponentially as $ k/(\xi a H) \rightarrow 0$. In contrast, the (--) mode is not excited out of its vacuum state, and hence remains with mode function
\be
A_{k-}\sim \frac{1}{\sqrt{2 k}} .
\ee

The resulting primordial spectrum of scalar anisotropies in V-mode polarization was worked out in \cite{Alexander:2017bxe}. The result, in the backreaction-limited regime $\xi \gg 1$ \cite{McDonough:2016xvu}, is
\bea \label{Vfluctamp}
\delta V_{k} = \frac{V(\phi_{end})}{ (2 \xi H)^{3/2}} \frac{f}{\alpha M_{Pl}} ,
\eea
which gives the primordial V-mode ($\Theta_V$) power spectrum,
\be
P_{V}(k) = \frac{1}{16 \pi^2 \xi^3} \left( \frac{f}{\alpha M_{Pl}}\right)^2 \left( \frac{k}{a H}\right)^3.
\ee
The scaling $P_V \sim k^3$ indicates a deep blue tilt, parametrized by the spectral index $n_V = 4$. This naturally suppresses the signal on large scales, which diminishes the possibility of observation.

The spectrum can be made scale invariant via a simple modification, which introduces a coupling of the axion to photon kinetic term, such that the gauge sector is given by
\be
\label{modmodel}
\mathcal{L} = I^2(\phi) \left( -\frac{1}{4}F_{\mu\nu}F^{\mu\nu} +   \frac{\alpha}{f} \phi F_{\mu\nu}\tilde{F}^{\mu\nu} \right) .
\ee 
This is similar to the magnetogenesis model of \cite{Caprini:2014mja}, which is equivalent to our model under the identification $\alpha \phi/f = constant =  \gamma $.

The gauge field evolution in this model depends sensitively on the time-dependence of $I(\phi)$. For example, for $V(\phi)= \frac{1}{2}m^2 \phi^2$ the correspondence between time and field dependence is given by,
\be
I(\tau) \sim a(\tau)^{n} \leftrightarrow I(\phi) = e^{- (n/4) \phi^2 / M_{Pl} ^2} .
\ee
The spectral tilt of the resulting electromagnetic spectra will depend on the value of $n$. In the absence of the axionic coupling, these are given by (see e.g.~\cite{BazrafshanMoghaddam:2017zgx,Caprini:2014mja})
\be
\label{tilts}
n_B = 5 - 2 |n  + \frac{1}{2}| \;\; , \;\; n_E = n_B - 2 ,
\ee
where we use the convention that scale invariance corresponds to $n_{B/E} = 0$. We will consider $n<0$, for which $I>1$ at all times, avoiding any period of strong coupling $e_{eff} = e/I \gtrsim1$. In this case, $I(\phi)$ is decreasing during inflation, and settles to $1$ at the end of inflation. This also generates a hierarchy between electric and magnetic fields on large scales, $B /E \ll 1$.

The analysis of \cite{Caprini:2014mja} confirmed that the model \eqref{modmodel} maintains the predictions for the spectral tilts \eqref{tilts}, while also maintaining the chiral asymmetry of the axion model \eqref{eq:axionmodel}.  In this case, one can straightforwardly compute the spectral tilt of V-mode anisotropies, which is a convolution of electric field anisotropies, similar to computation of the induced curvature perturbations \cite{McDonough:2016xvu}.  This leads to the relation,
\be
n_V = n_E +1 ,
\ee
and thus scale-invariant $V$-modes, $n_V=1$, occur for scale-invariant electric fields $n_E=0$. The coupling function is thus $I(\phi)\sim a^{n}$ with $n=-2$.

We can see this in more detail as follows. The gauge field equation of motion is given by,
\begin{equation} \label{EoMA}
\frac{d^2 \tilde{A}_{k\pm}}{d \tau^2} + \bigg[  k^2 \mp \, 2 \frac{k}{\tau} \frac{\alpha}{f}\left( n \phi - \sqrt{2 \epsilon} M_{Pl} \right) - \frac{n(n+1)}{\tau^2} \bigg] \tilde{A}_{k\pm} \, = \, 0 \, ,
\end{equation}
where $\tilde{A}$ is the re-scaled variable,
\be
\tilde{A} = I A. 
\ee
During slow-roll inflation, the second term in round brackets is sub-dominant, and the equation of motion can be expressed as
\be
\label{Aeom}
\frac{d^2 \tilde{A}_{k\pm}}{d \tau^2} + \bigg[  k^2 \pm\, 2 k \frac{\hat{\xi} }{\tau} - \frac{n(n+1)}{\tau^2} \bigg] \tilde{A}_{k\pm} \, = \, 0  ,
\ee
where we define the parameter $\hat{\xi}$ as ,
\be
\hat{\xi} = \frac{\alpha}{f} |n| \phi ,
\ee
where again we note that we are in the regime $n<0$. During slow-roll inflation $\phi$ is approximately constant, and $\hat{\xi}$ can be treated in an adiabatic limit.

The equation of motion \eqref{Aeom} displays three distinct regimes. At very early times, the first term in square brackets dominates, and the gauge field is in the Bunch-Davies vacuum. At late times, the third term dominates and the standard magnetogensis analysis of electric and magnetic takes over. However, for an intermediate regime,
\be
  \frac{|n(n+1)|}{2 \hat{\xi}}< |k\tau| < 2 \hat{\xi} ,
\ee
the system displays the same tachyonic instability as the axion model \eqref{eq:axionmodel}. The resulting behavior on large scales $| k\tau | \ll 1/\hat{\xi}$ is given by \cite{Caprini:2014mja}
\bea
&& A_+ \simeq \sqrt{\frac{- \tau}{2 \pi}} e^{\pi \hat{\xi}} \Gamma( |2 n+1 |) |2 \hat{\xi} k \tau |^{- |n+ 1/2|}  ,\\
&& A_- \simeq 0 .
\eea
From which one can compute the spectral indices \eqref{tilts}. As in the minimal axion model \cite{Alexander:2017bxe}, the amplitude of $P_V$ is enhanced by the exponential factor $e^{2 \pi \hat{\xi}}$, %

Ignoring backreaction considerations, the power spectrum of V-mode anisotropies is given by
\be
P_{\Theta_V} \simeq \frac{(\hat{\xi} a H)^4}{M_{Pl} ^4} \frac{e^{4 \pi \hat{\xi}}}{\hat{\xi}^6} .
\ee
The observational predictions are succinctly characterized by the circular-polarization-to-scalar ratio $r_V$ and spectral index $n_V$, given by
\be
r_V = \varepsilon \left( \frac{H}{M_{Pl}}\right)^2 \frac{e^{4 \pi \hat{\xi}}}{\hat{\xi}^2}\; \;\; ,\;\;\;  n_V = 1. 
\ee
where $\varepsilon$ is the inflationary slow-roll parameter. This can easily give $r_V = \mathcal{O}(1)$, e.g.~for $H=10^{-10} M_{Pl}$ one finds $r_V=1$ for $\xi=4.2$.

%=================================================================%
\section{Boltzmann Hierarchy for $V$-mode Polarization}
\label{sec:CMB}
%=================================================================%

We have established the circular polarization can be produced during inflation, and computed spectrum of scalar anisotropies. We now consider the propagation of this from the end of inflation (or reheating) to last scattering.  

\subsection{On Primordial Electric Fields}

In the absence of scatterings, any initial $V/I$ propagates through the universe unchanged. This occurs because both $V$ and $I$ redshift as $a^{-4}$ and hence the circular polarization fraction $V/I$ is unaffected by the expansion of the universe. On the other hand, with scatterings included, Thomson scattering exponentially diminishes circular polarization. This can be understood both from the Boltzmann equation for $V$ and from the equation of motion describing large scale fluctuations of the gauge field $A_{\mu}$. This latter has been studied extensively in the context of primordial magnetic fields e.g.~\cite{Martin:2007ue,Subramanian:2015lua}. For our purposes, we are interested in primordial \emph{electric} fields.

The action for the gauge field $A_{\mu}$ during the radiation dominated era is \cite{Martin:2007ue},
\be
S[A_{\mu}] = \int {\rm d}^4 x \sqrt{-g} \left( \frac{1}{4}F^2 - j^\mu A_{\mu}\right)
\ee
The current $j^\mu$ is given by Ohm's law,
\be
j^{\mu} = \sigma E^{\mu},
\ee
where $\sigma$ is the conductivity of the radiation plasma, which scales with the number density of electrons $\sigma \propto n_e$.

The resulting equation of motion for Fourier modes the gauge field is given by,
\be
\ddot{A}_{k\pm} + \left( H + \sigma \right)\dot{A}_{k\pm} + k^2 A_{k\pm} =0 .
\ee
In the limit of large scales $k/aH\rightarrow0$ and high conductivity $\sigma \gg H$, the equation of motion is simply
\be
\ddot{A}_{k\pm} + \sigma \dot{A}_{k\pm} =0 ,
\ee
which leads to the solution
\be
A_{k\pm} = \frac{D_{1\pm}(k)}{\sigma} e^{- \sigma t} + D_{2\pm}(k) ,
\ee
and hence an exponential suppression of both $A_{+}$ and $A_-$. Assuming an initial condition $A_+ \gg A_-$,  $V(\eta=0)=V_0$, this gives the circular polarization $V$ as
\be
V \simeq \dot{A}_{+}^2 = V_0 e^{- 2 \sigma t} .
\ee
Thus any initial super-Horizon circular polarization $V_{0}$ is exponentially suppressed by the finite density of electrons in the radiation dominated phase.

\subsection{The Boltzmann Hierarchy for V}

We now return to CMB physics. The Boltzmann equation for scalar brightness perturbations in V is given by \cite{Kosowsky:1994cy},
\be
{V} '  + k \mu V + \dot{\tau} V= \frac{3}{2}\dot{\tau} V_1 .
\ee
Following the textbook treatment of CMB anisotropies, this admits an integral solution given by
\be
\label{solution}
V_\ell = \int_0^\eta \frac{3}{2}\dot{\tau}e^{-\tau}V_1(\eta') \, j_\ell'(k(\eta-\eta')) {\rm d} \eta' ,
\ee
which in the sudden decoupling approximation, $\dot{\tau}e^{-\tau} =\delta(\eta' - \eta_{\rm l.s.})$,  is simply,
\be 
\label{Vell}
V_\ell \simeq \frac{3}{2}V_1(\eta_{\rm l.s.})j'_\ell(k(\eta-\eta_{\rm l.s. })).
\ee
The resulting $V_{\ell}$ thus depend on the value of $V_{1}$ at last scattering, which is dictated by the first moments of the Boltzmann hierarchy,
\bea
&& V_0 ' + \dot{\tau}V_0 = - k  V_1 \\
&& V_1 ' + \frac{1}{2}\dot{\tau}V_1 = -\frac{2}{3} k  V_2 + \frac{1}{3} k V_0 
\eea
and a similar equation for $V_2$.

The distinction between circular polarization anisotropies and temperature anisotropies can be appreciated by studying the second order differential equations for $V_0$ and $V_1$, in the limit that $\dot{\tau}$ is constant,
\bea
&& V_0 '' + \dot{\tau}V_0 ' - \frac{1}{3} k^2 V_0= \frac{1}{2} k \dot{\tau} V_1 , \\
&& V_1 '' + \dot{\tau}V_1 ' + \frac{1}{3} k^2 V_1= - \frac{1}{3} k \dot{\tau} V_0 .
\eea
This system does not admit a undamped oscillatory solutions, but instead both $V_0$ and $V_1$ inherit an exponential suppression from the friction term $\dot{\tau}V_{0,1}'$. This corrects an error in \cite{Giovannini:2010ar}, where the term $\dot{\tau}V_{0,1}'$ was dropped in going from the first to second order equations of motion for from $V_{0,1}$, and this exponential suppression was missed.

To solve for $V_{0,1}$, we follow \cite{Hu:1996mn}, and work in a series expansion in $k/\dot{\tau} \ll 1$. Before recombination one has $k/\dot{\tau} \ll 1$ for all modes observable in the CMB (see e.g.~\cite{Hu:1996mn}).  This hierarchy is fundamental to the study of CMB physics: it causes an exponential suppression of $\ell \geq 2$ temperature multipoles, preventing power from being transferred out of the monopole and dipole, which allows the universe to be treated as a photon-baryon fluid described solely for $\ell=0,1$.

Solving for $V_0$ at zeroth order in $k/\dot{\tau} \ll 1$ gives,
\be
V_0 (\eta)= V_0 (0) e^{- \int _{0} ^{\eta} \dot{\tau} \mathrm{d} \eta'},
\ee
and similarly for $V_1$,
\be
V_1 (\eta)= V_1 (0) e^{- \frac{1}{2}\int _{0} ^{\eta} \dot{\tau} \mathrm{d} \eta'}.
\ee
 where we apply initial conditions at $\eta=0$, or equivalently, at reheating.
 
It is also important to note the definition of the optical depth $\tau$,
 \be
 \tau = \int _{\eta} ^{\eta_0} \dot{\tau}\, \mathrm{d}\eta'.
 \ee
Since $\dot{\tau}$ is decreasing with time in an expanding universe, and $\tau$ is defined as an integral from the past to today, the integral of $\dot{\tau}$ is dominated by $\dot{\tau}$ at the first moments More precisely, for $\eta \gg 0$,
\be
  \int _{0} ^{\eta}\dot{\tau} \rm{d}\eta' = \int_{0} ^{\eta_0} \dot{\tau} \mathrm{d}\eta' - \int _{\eta} ^{\eta_0} \dot{\tau} \mathrm{d}\eta' \equiv \tau_0 - \tau_{\rm \eta}  \simeq \tau_0
\ee
where $\tau_0$ is the optical depth at the beginning of the radiation dominated era.

From this it follows that,
\be
V_0 (\eta)\simeq V_0 (0) e^{- \tau_0 },
\ee
and similarly for $V_1$,
\be
V_1 (\eta) \simeq V_1 (0) e^{- \frac{1}{2}\tau_0 },
\ee
where $\tau_0$ is defined as the optical depth at the beginning of the radiation dominated era.

We can now compute the $\mathcal{O}(k/\dot{\tau})$ correction to $V_1$, which captures the $V_1$ sourced by $V_0$. This is given by,
\be
\label{V01sol}
V_{1sourced}(\eta)  = \frac{k \eta }{3} e^{- \tau_0} V_0 (0).
\ee
Evaluating this at last scattering determines the $V_{\ell}$ via equation \eqref{Vell}.

\subsection{Primordial V-Modes in the CMB}

The correlation function $C_{\ell} ^{VV}$ can now be straightforwardly evaluated in terms of the primordial power spectrum. The general form of the solution is given by,
\be
C_\ell^{VV} (\eta)  =   \frac{9}{4}\int _{0} ^{\infty} \frac{k^3}{2 \pi^2} |V_1 (\eta_{\rm l.s.})|^2  j'_\ell(k(\eta-\eta_{\rm l.s. }))^2 {\rm d}\log k. \\
 \ee
Using equation \eqref{solution} and\eqref{V01sol}, we have
\be
C_\ell^{VV} =   \frac{1}{4}\int _{0} ^{\infty} \frac{k^2 \eta_{\rm l.s}^2}{9} e^{- 2 \tau_0 } P_{V}(k)  j'_\ell(k(\eta-\eta_{\rm l.s. }))^2  {\rm d}\log k,
\ee
where we have used equation \eqref{V01sol} to relate $|V_1|^2$ to the primordial power spectrum \eqref{PVkV0}. 

To estimate this, we follow the canonical estimation methods of CMB physics \cite{Mukhanov:2003xr}. We consider that $j_{\ell} ' (z) ^2$ is roughly peaked at $z\simeq 2 \ell$, which is strictly true only of $\ell \lesssim 30$, in order to pull terms out of the integral, and we evaluate the Bessel function integral exactly,
\be
\int_{0} ^{\infty} d\log x\; j_{\ell}'(x)^2 = \frac{1}{6} \frac{1}{\ell (\ell +1) - 2} .
\ee
This leads to
\be
\ell (\ell +1)C_\ell^{VV} (\eta_0)  \simeq \frac{1}{24} \ell^2  \left( \frac{\eta_{\rm l.s.}}{\eta_{0}} \right)^2 e^{- 2 \tau_0}   P_{V}(k) .
\ee
We can further approximate $\eta_{\rm l.s.}/\eta_0 \sim 1/30$ which gives for $\ell \simeq 10$,
\be
\ell (\ell +1)C_\ell^{VV} (\eta_0) \simeq 10^{-3} e^{- 2 \tau_0}   P_{V}(k).
\ee
Thus in addition to the $10^{-3}$ suppression that comes simply from the $k/aH$ scaling of the source, there is an additional exponential suppression from the optical depth at the beginning of the radiation dominated era.

\begin{figure}
\vspace{1cm}
\includegraphics[scale=0.4]{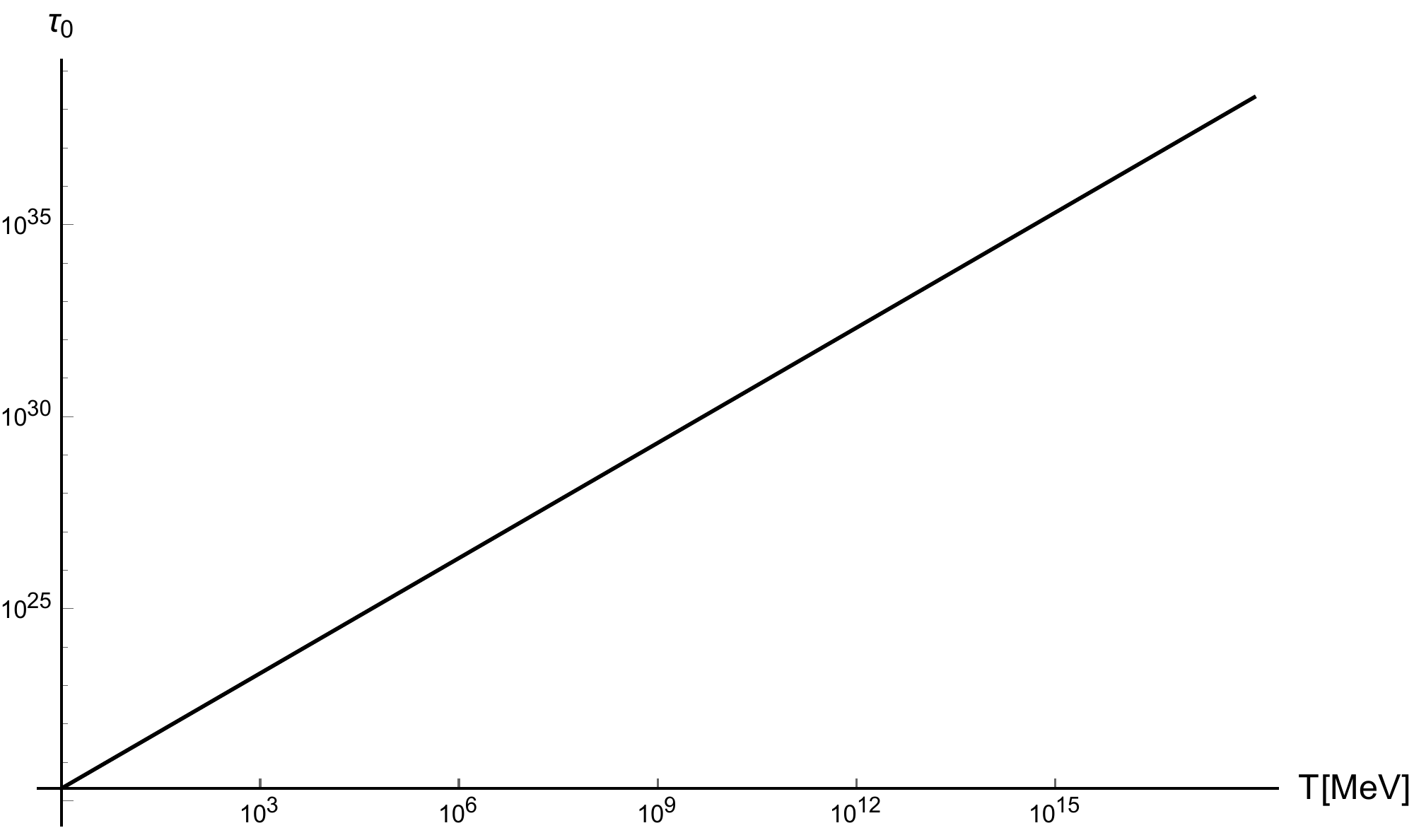}
\caption{Optical Depth at the beginning of radiation domination as a function of the reheating temperature (in units of MeV).}
\label{OpticalDepth}
\end{figure}

The numerical value of this suppression factor can be computed using a CMB solver, such as CLASS, which natively computes $\tau$ up to a redshift of $z\simeq 10^5$. At higher redshifts, one may calculate the evolution of $\tau(z)$ analytically. The precise value of $\tau_0$ is determined by the temperature of the universe at the beginning of the radiation dominated era, or equivalently, the reheat temperature. Utilizing in this way the mathematica package CMBquick \cite{CMBquick}, we find
\be
\tau_0 \gtrsim 10^{20} ,
\ee
where the lower bound corresponds to the lower bound for the reheat temperature,  $T_{re} \simeq \mbox{MeV}$. A plot of $\tau_0$ as a function of temperature $T$ is given in Figure \ref{OpticalDepth}.

It follows that the numerical value for $\ell (\ell +1)C_\ell^{VV}$ is given by,
\be
\frac{\ell (\ell +1)}{2 \pi}C_\ell^{VV} (\eta_0)  \simeq 10^{- 10^{20}} P_{V}(k).
\ee
Expressed in units of temperature and the circular-to-scalar ratio $r_V$, this gives 
\be
\frac{\ell (\ell +1)}{2 \pi}C_\ell^{VV} (\eta_0)\sim 10^{- 10^{20}} r_V \mu {\rm K}^2.
\ee
which is well below any observable prospects.

%=================================================================%
\section{Discussion: Gravitational Waves and Other Sources of Primordial $V$}
\label{sec:conclusion}

%=================================================================%

In this work we have studied the generation of primordial circular polarization, and the imprint on the cosmic microwave background. The early universe production of $V$ is a generic consequence of axion couplings to gauge fields during inflation, and gives rise to a model-dependent spectral tilt of anisotropies. However, while $V$ can indeed be produced in large quantities during inflation, we have demonstrated the relic $V$ remaining in the CMB is negligibly small, suppressed relative to the primordial $V$ by a factor of $\gtrsim 10^{10^{20}}$, and hence is unobservable in any conceivable future experiment.  

However, the direct production of $V$ is not the only mechanism for sourcing $V$ from primordial perturbations. One alternative is to consider the $V$-anisotropies induced by chiral primordial gravitational waves.  The connection between these can be appreciated by considering the anomaly for the electron axial vector current (see e.g.~\cite{Harvey:2005it}),
\be
\label{anomaly}
\partial_\mu j^{\mu 5} = \frac{e^2}{8 \pi^2} F \tilde{F} +  \frac{1}{384 \pi^2}R \tilde{R}  .
\ee
The first term on the right hand side is simply $F \tilde{F} = \dot{A}_L \cdot k A_{L}  - \dot{A}_R \cdot k A_{R}$, which is of course crucial to the chiral asymmetry produced during inflation, while the second term can expressed in terms of helicity modes of gravitational waves \cite{Caldwell:2017chz},
\bea
\label{eq:GWs}
&&  R \tilde{R} =   \int \mathrm{d}\log k \left[ k^3 (\Delta_R^2 - \Delta_L ^2) - k ({ \Delta_R '} ^2 - {\Delta_L '} ^2)\right]  \nonumber \\
&&\Delta_P ^2 = \frac{k^3}{\pi^2} |h_{P,k}|^2 \;\;,\;\;{ \Delta_P'} ^2 = \frac{k^3}{\pi^2}  |{h'_{P,k}}|^2  ,
\eea
where $P=L,R$ denotes the handedness, which is non-zero if the gravitational wave spectrum is circularly polarized, that is, $h_{L} \neq h_{R}$. This connection is particularly interesting considering the role of chiral gravitational waves in leptogenesis \cite{Caldwell:2017chz,Alexander:2004us}, superfluid dark matter \cite{Alexander:2018fjp}, and string theory\cite{McDonough:2018xzh}.

In the framework of cosmological perturbation theory, this arises as a second order effect, similar to the production of $V$ due to photon-photon scattering \cite{Motie:2011az,Sawyer:2014maa,*Sawyer:2012gn,Sadegh:2017rnr,Montero-Camacho:2018vgs,Kamionkowski:2018syl,Inomata:2018rin}. The relevant scattering process in this case is \emph{graviton-photon} scattering \cite{Bjerrum-Bohr:2014lea} . One naturally expects this process to be Planck suppressed, and scale with the tensor-to-scalar ratio $r$, and but it may nonetheless lead to a signal that is orders or magnitude larger then those found thus far. We leave a precise evaluation of the induced $V$ to future work.

An alternative possibility is to consider new interactions of photons, that contribute new collision terms to the Boltzmann equation (as in e.g.~\cite{Alexander:2008fp,Bavarsad:2009hm,Batebi:2016ocb,Ejlli:2017uli}) or alter the propagation the photons. Both of these arise in axion quintessence scenarios, with axion-photon scattering contributing a new collision term and the axion velocity $\dot{\phi}$ altering the photon dispersion relation. Finally, we also note that we have not endeavored to study the vector or tensor Boltzmann equations for $V$, and their contribution to the $V$-mode power spectrum. This will appear in a companion paper \cite{companion}, along with a comprehensive analysis of additional source terms to the Boltzmann equation and their observational consequences.

\vspace{0mm}
\acknowledgments
The authors thank Yacine Ali-Ha\"{i}moud, Robert Brandenberger, David Kaiser,  Alan Guth, Wayne Hu, Anthony Pullen, Bradley Shapiro, David Spergel, and Kendrick Smith,  for insightful comments and suggestions. EM is supported in part by the National Science and Engineering Research Council of Canada via a PDF fellowship.

\bibliography{VmodeRefs}

\bibliographystyle{utphys}

\end{document}